\documentstyle[twocolumn,aps,prl]{revtex}            
\draft
\begin{document}

\twocolumn[\hsize\textwidth\columnwidth\hsize\csname@twocolumnfalse\endcsname 

\begin{flushright}
{\tt to appear in Phys. Rev. B (2000)} 
\end{flushright}

\title{Anomalous Sliding Friction and Peak Effect near the Flux
Lattice Melting Transition}

\author{E. Granato}
\address{Laborat\'orio Associado de Sensores e Materiais,
Instituto Nacional de Pesquisas Espaciais, \\
12201-190 S\~ao Jos\'e dos Campos, S\~ao Paula, Brazil}

\author{T. Ala-Nissila}
\address{Helsinki Institute of Physics and Laboratory of Physics,\\
Helsinki University of Technology, P.O. Box 1100,\\
FIN-02015 HUT, Espoo, Finland}

\author{S.C. Ying}
\address{Department of Physics,
Brown University, \\
Providence, Rhode Island 02912}

\maketitle

\begin{abstract}

Recent experiments have revealed a giant "peak effect" in
ultrapure high $T_c$ superconductors. Moreover, the new data show
that the peak effect coincides exactly with the melting
transition of the underlying flux lattice. In this work, we show
using dynamical scaling arguments that the friction due to the
pinning centers acting on the flux lattice develops a singularity
near a continuous phase transition and can diverge for many
systems. The magnitude of the nonlinear sliding friction of the
flux lattice scales with this atomistic friction. Thus, the
nonlinear conductance should diverge for a true continuous
transition in the flux lattice or peak at a weakly first order
transition or for systems of finite size.

\end{abstract}


]

One of the central unsolved problems in type-II superconductivity
concerns the so-called "peak effect". When a current $I$ is
passed through the superconductor in the mixed phase, the flux
line lattice (FLL) moves in response to the Lorentz force,
leading to dissipation and an induced voltage $V$. Naively, the
nonlinear conductance $C=I/V$ is expected to decrease
monotonically towards the $H_{c2}$ phase boundary because of the
diminishing order parameter and hence a reduced pinning strength.
However, experimentally it was observed a long time ago that
instead of a monotonic behavior, the conductance  peaks sharply
to a large value before dropping  at the superconducting-normal
transition\cite{leblanc,bhl,kes,marley}. It has also been
established that this peak effect is not just limited to
conventional superconductors, but shows up in a similar fashion
in the high $T_c$ YBCO superconductors also\cite{ling91,kwok}. To
date, however, there has been no satisfactory explanation for
this peak effect, although various possible mechanisms have been
proposed as the origin of this phenomenon. One popular idea based
on the collective pinning theory\cite{lo} is that the FLL softens
towards the $H_{c2}$ boundary, leading to a smaller elastic coherence
length (Larkin length) and enhanced stronger pinning by the
impurities\cite{lmv}. It is not clear though how this mechanism
can give rise to the sharp peak in the conductance. Moreover,
recent data have revealed giant peak effects in ultrapure high
$T_c$ superconductors with as much as a 35-fold increase in $C$
from onset to peak\cite{ling99}, which is hard to explain with
the collective pinning idea. It has also been widely suggested
that the peak effect is either caused by or related to an
underlying FLL melting transition\cite{kwok,lmv,bh,tang}. Up
until very recently, this idea has remained speculative because
of the difficulty of a direct experimental observation of the FLL
melting transition. Two recent studies have now established
conclusively the relation of the peak effect with the underlying
phase transition in the FLL\cite{ling99,ling00}. In the
study\cite{ling99} involving  ultrapure YBCO, the peak effect is
shown to coincide exactly with the point at which there is a small
magnetization jump  $\Delta M$ or discontinuity in the slope of
the magnetization. This behavior of $\Delta M$ is interpreted as
the signature of either a continuous or very weak first order
transition. In another study of the conventional
superconductor\cite{ling00} Nb, an AC magnetic susceptibility
($\chi$) measurement was made in conjunction with a small angle
neutron scattering (SANS) study of the structure of the
underlying FLL. The peak effect (a dip in the real part of the
AC susceptibility $\chi$)
is observed to occur precisely at the point where the diffraction
peak in the SANS pattern of the FLL begins to broaden into
ring-like features. In this case, the transition is more strongly
first order, with a direct observation of hysteresis involving
superheating and  supercooling behavior.

We have, in an earlier paper\cite{agy}, suggested that the
sliding friction for the FLL would be anomalously large near a
continuous or very weak first order melting transition due to the
enhanced coupling of the pinning centers to the FLL through the
critical fluctuations. The central idea is that the mobility of
the FLL is not just controlled by the pinning strength of the
impurities, which is an equilibrium property. It depends also on
the non-adiabatic coupling of the pinning center to the dynamical
excitations of the FLL, leading to a frictional damping $\eta$ on
the FLL. Recent theoretical developments in understanding
nonlinear sliding friction of an adsorbed
monolayer\cite{persson,robbins,tomassone,gy} in the boundary
lubrication problem are particularly helpful in elucidating this
problem. Aside from inertia mass effects, the behavior of these
systems is very similar. In the FLL, the driving force $F$ is the
Lorentz force proportional to the current passing through the
superconductor, and the moving FLL produces a changing flux and
an induced voltage that is proportional to the average drift
velocity $\langle v \rangle $ of the FLL. Thus, in the language
of the boundary lubrication problem, the static friction of the
adlayer corresponds to the critical current $J_c$ in the
superconductor, and the nonlinear sliding friction $\bar{\eta}$
of the adlayer defined as $\bar{\eta}=F/\langle v \rangle$ is
essentially the nonlinear conductance $C$ for the superconductor.
Note that $\bar{\eta}$ is just the inverse of the usual
definition of the mobility for the adlayer. In the discussion of
the peak effect problem, the nonlinear conductance related to the
sliding friction $\bar{\eta}$ of the FLL is actually the more
relevant quantity. Near the occurrence of the peak effect, the
$I-V$ curve is often of such a nature that there is a continuous
rise of the voltage with increasing driving current such that the
exact value of threshold critical current $J_c$ is ill-defined,
and the nonlinear conductance is a better measure of the anomaly
for the mechanical response of the FLL\cite{anderson}. In the AC
magnetic susceptibility measurement, what determines the
magnitude of the screening current and hence the magnitude of the
susceptibility is clearly the nonlinear conductance and not so
much a single threshold critical current density $J_c$. Results
from various numerical studies of the boundary layer
problem\cite{persson,robbins,gy} have shown that both the static
friction and the nonlinear sliding friction depend in a
complicated manner on the interplay of the strength of the
pinning potential, interactions among the particles (vortices for
the FLL) and the bare frictional damping $\eta$ from the
environment. In this paper, we will quantify the concept that for
the FLL, it is the variation of the nonadiabatic frictional
damping $\eta$ and not the adiabatic pinning strength that
develops anomalous temperature and magnetic field dependence near
the melting transition. This anomalous behavior of the frictional
damping then leads to the peak effect for the nonlinear sliding
friction $\bar{\eta}$ for the FLL and hence the conductance C of
the superconductor.  We show below through general dynamical
scaling arguments the explicit singularity of the friction $\eta$
near the transition.

Let us first consider the random force acting on the pinning
center at the position ${\bf r}$ by the flux lattice. In the
simple pair interaction model, this can be expressed in terms of
the linear displacement $u_{{\bf q}, \alpha}$ of each vortex from
its equilibrium position as

\begin{equation}
f_{\alpha}({\bf r})=\sum_{{\bf q}}W({\bf r},{\bf q} )u_{{\bf q},
\alpha}. \label{rforce}
\end{equation}
Here, ${\bf q}$ stands for the normal mode index of the FLL,
$\alpha$ is the Cartesian component label and $W$ represents the
coupling function. In response to this, there is an equal and
opposite reaction force on the FLL by the pinning center. When
correlations between the different pinning centers are neglected,
the frictional damping (in the Markovian limit) on the FLL is
then given by\cite{mori,ying}

\begin{equation}\eta=\sum_{{\bf q},{\bf r} }W^{2}({\bf r},
{\bf q})S({\bf q}, \omega=0), \label{fric}
\end{equation}
where $S({\bf q},\omega=0)$ is the dynamic structure factor
defined as $\int^{\infty}_{0}\,dt \langle u_{{\bf q},
\alpha}(t)^* u_{{\bf q}, \alpha}(0) \rangle $. Correlations
between the random forces from pinning centers at different
positions would lead to higher order terms in the pinning center
concentration $n_p$ in Eq.(\ref{fric}), and are negligible in the
limit $n_p \rightarrow 0$. According to general dynamical scaling
arguments \cite{halperin69,halperin77}, $S({\bf q}, \omega)$ should
take the scaling form  near $T_c$ for a continuous phase
transition as

\begin{equation}
N^d S({\bf q},\omega )=\xi ^{z+\gamma /\nu }g_{\pm }(q\xi, \omega
\xi ^{z})\text{ ,} \label{scal}
\end{equation}
where $g_{\pm }$ is a scaling function, $\xi \propto \vert
T/T_{c}-1 \vert^{-\nu}$ is the divergent correlation length,  $d$
is the system dimension, $\gamma$ the susceptibility exponent and
$z$ is the dynamical critical exponent. Substitution of Eq.
(\ref{scal}) back into Eq. (\ref{fric}) then leads to the
conclusion that as one approaches $T_c$, the friction $\eta$ has
a singular part that goes as $\eta \propto \vert T/T_{c}-1
\vert^{-x}$ with $x=\nu(z-d)+\gamma$. The dimension $d$ enters
explicitly through the $q$ integration in Eq. (\ref{fric}) where
we have assumed a typical short ranged coupling potential $W(q)$
that is regular at $q=0$. Thus the friction $\eta$
can either diverge if $x>0$ or be finite with a cusp only \cite{anomaly}. 
Similar anomaly has also been predicted for adatom diffusion near the
surface reconstruction transition of the W(100) surface
\cite{ahy}. For this case, the exponent $x$ has been explicitly
evaluated for a model Hamiltonian and shown to have the
value \cite{ahy,gy00} $x \approx 1.8$. Thus the diffusion constant
of adatoms on this surface is predicted to vanish at the
transition.

The friction $\eta$ calculated in Eq. (\ref {fric}) corresponds
to the bare friction acting on the center of mass (CM) degree of
freedom of the FLL. It is analogous to the friction exerted by
the substrate on an adsorbed layer in the boundary lubrication
problem. Experimentally, the mobility measurements of the FLL have
all been performed in the nonlinear regime. In the presence of an
external pinning potential, the CM motion of the FLL is coupled
to the single vortex motion which depends in turn on the
interactions with other vortices. Thus the nonlinear response of
the flux lattice under a driving current can only be determined
by solving the coupled  Langevin equations.  In general, the
nonlinear sliding friction  $\bar{\eta}$  of the FLL depends on
the details of the vortex interaction, strength of the pinning
potentials, and the driving force. However, in various recent
studies of the nonlinear sliding friction of an adsorbed
overlayer on an substrate\cite{persson,gy}, it has been shown
that the magnitude of $\bar{\eta}$ is determined by the bare
friction $\eta$ as given in Eq.(\ref{fric}), with $\bar{\eta}$
approaching the bare friction $\eta$ in the limit of large
driving force. Even for a system with a positive exponent $x$
leading to a divergent behavior for $\eta$ and $\bar{\eta}$ near
the transition, the conductance peak at the transition in practice
will be significantly rounded by crossover effects due to the
nonzero driving current. It has been argued, in general terms,
that the current density $J$ sets an additional length scale in
resistance measurements\cite{huse} $L_J^{d-1} \sim kT/J$ due to
thermal fluctuations. The divergent critical fluctuations at
$T_c$ will be then be cut off by this length when $\xi\sim L_J$,
giving rise to a nonlinear resistance behavior $R\sim
I^{x/\nu(d-1)}$. Experimentally \cite{ling99,bh}, a strong nonlinearity is indeed
observed for the conductance maxima which decreases for increasing
$I$. Thus, we conclude that for a FLL system with a positive
exponent $x$, its nonlinear sliding friction $\bar{\eta}$ has a
peak at the melting transition, its origin being the strong
critical fluctuation near the melting transition. This then
leads to the peak in the conductance C. In the case of a weak
first order transition or finite size system, the divergence or
the cusp singularity of $\eta$ would be rounded off even in the
linear regime and thus we expect the peak effect for these
systems to be much weaker.

Now we come back to the recent experimental data on peak effect
and discuss them in light of the above theoretical considerations.
Much of the difficulty associated with understanding the FLL
dynamics starts with the fact that we do not even have a very
detailed understanding of the ground state. The accepted picture
now for the weak pinning limit is that of a Bragg glass with
quasi-translational long range order (LRO) and true orientational
LRO\cite{giamarchi}, due to the presence of random pinning
centers. Similarly, we do not have a clear picture of how or
whether the FLL melts just before the superconducting-normal
transition\cite{nelson,houghton,huse}. The recent
data\cite{ling99,ling00} strongly support the existence of a
phase transition in the FLL just before the $H_{c2}$ phase
boundary. For an ultrapure sample of the high $T_c$ YBCO
superconductor, static magnetization measurements show a very
small discontinuity at high magnetic field (5 T) and no
discernible jump but only a discontinuity in the slope of the
magnetization at lower fields. This is identified as the melting
transition, the transition being continuous at low fields and
weakly first order at high fields. The "peak effect", identified
by the dip in the real part of the AC magnetic susceptibility
$\chi$ occurs at precisely the same temperature and magnetic
field as this "melting" transition. This peak effect is
"gigantic" involving  a 35-fold increase in the nonlinear
conductance $C$ through a narrow range of change of temperature
or the magnetic field. This is much stronger than all the
previously observed peak effects which typically show a peak to
onset ratio of 3 to 4. According to the present theory, this
sharp peak behavior in $C$ can be understood as arising from the
sharp rise in the friction acting on the FLL due to the coupling
of the pinning centers to the strong critical fluctuations near
the continuous or weakly first order melting transition.
According to our scaling arguments, the existence of a peak
effect require that the exponent $x=\nu(z-d)+\gamma$ be positive.
At the moment, there exists no detailed information on any of
these exponents for the FLL melting transition in the presence of
pinning centers. However, existing calculations of the dynamical
exponent $z$ for disordered systems \cite{young} give results which are
generally larger than $z=4$. Thus, it is entirely
plausible that the corresponding exponent $x$ for the FLL can be
positive. In practice, the divergence of the critical
fluctuations will be cut off by the length scale set by the
current $L_J^{d-1} \sim kT/J$ . In addition, imperfections in the
crystalline order of the sample also provides a cutoff. This
explains then the gigantic peak effect for the ultrapure YBCO as
opposed to the much smaller peak effect for the poorer quality
samples. Another feature of the data that supports the present
theory is the large width of the $\chi'(T)$ dip. At $H=5.0$ T,
the width of the $\chi'(T)$ dip is about 1 K while the width of
the $\Delta M$ discontinuity is only about\cite{ling99} $0.08$ K.
This can be understood from the fact that  $\chi'(T)$ measures
the critical fluctuations through its dependence on the friction
while $\Delta M$ is just an order parameter measurement connected
with the density of the vortices. In addition to this study for
YBCO, there is also a recent study for the conventional
superconductor Nb involving simultaneous small angle neutron
scattering (SANS) as well as AC magnetic susceptibility
measurements \cite{ling00}. The melting transition here can be clearly
identified as the point where the sharp Bragg-like peak in the
ordered FLL phase first begins to broaden into ring-like
features. By contrast with the high $T_c$ YBCO material, the
stronger first order nature of the melting transition in Nb is
clearly evidenced by the observation of superheating and
supercooling below and above the melting transition
\cite{ling00}. Again, the peak effect as determined from the
magnetic susceptibility measurements coincides with the melting
transition. However, the peak effect in this case is much weaker,
and the conductance $C$ only shows a four-fold increase from
onset to the peak value. Since the transition here is of first
order, the critical fluctuations are much weaker and the
correlation length does not diverge at the melting transition
point. In fact, the situation here is similar to the poorer
quality sample of YBCO where impurities and imperfections cut off
the divergent critical fluctuations. As a result, the friction
acting on the FLL has only a weak maximum instead of a divergent
behavior at the transition point, and the corresponding peak
effect is much weaker.

In conclusion, we have presented here a general scaling argument
that the frictional damping  exerted by the pinning centers on the
flux lattice has a singularity (or a cusp) near a continuous
melting transition in the lattice. While most previous theoretical
considerations of the peak effect focus on the adiabatic pinning
strength, the present work identifies the origin of the peak
effect through the non-adiabatic coupling of the pinning centers
to the strong critical fluctuations near the transition point.
This leads to a vanishing  linear mobility for the flux lattice
at the transition. In the nonlinear regime, the finite driving
current provides a cutoff for the divergent critical
fluctuations, and this leads to  a finite peak in the nonlinear
sliding friction for the FLL and hence the conductance for the
superconductor, with the strength of the peak dependent on the
magnitude of the driving current. The recently observed gigantic
peak effect in high $T_c$ superconductors and the strong
correlation between the peak effect and the observed melting
transition provide strong support for the  mechanism proposed
here.

\bigskip
We acknowledge helpful discussions with X.S. Ling and Martin
Grant. This work was supported by in part by a joint NSF-CNPq
grant, by FAPESP (grant 99/02532-0) (E.G.), and the Academy of
Finland through its Center of Excellence program (T.A-N.).


\begin{thebibliography}{100}

\bibitem{leblanc}
M. A. LeBlanc and  W. A. Little in: {\it Proceedings of
Conference on Low Temperature Physics}, (University of Toronto
Press, Toronto), p.198 (1960).

\bibitem{bhl}
T. G. Berlincourt, R. R. Hake and  D. H. Leslie, Phys. Rev. Lett.
{\bf 6}, 671 (1961); S. H. Autler, E. S.  Roenblumand  K. Gooen,
Phys. Rev. Lett. {\bf 9}, 489, (1962); W. Desorbo, Rev. Mod.
Phys. {\bf 36}, 90 (1964).

\bibitem{kes}
P. H. Kes and C. C. Tsuei, Phys. Rev. B{\bf 28}, 5126 (1983).

\bibitem{marley}
A. C. Marley ,  M. J. Higgins and  S. Bhattacharya, Phys. Rev.
Lett. {\bf 74}, 3029 (1995).

\bibitem{ling91}
X. S. Ling and  J. I. Budnick, in {\it Magnetic Susceptibility of
Superconductots and Other Spin Systems}, edited by R. A. Hein et
al., Plenum Press, New York, N.Y., p.377 (1991).

\bibitem{kwok}
W. K. Kwok,  J. A. Fendrich,  C. J. Van der Beek and  G. W.
Crabtree, Phys. Rev. Lett. {\bf 73}, 2614 (1994).

\bibitem{lo}
A. I. Larkin and Yu. N. Ovchinnikov, J. Low Temp. Phys. {\bf 34},
409 (1979).

\bibitem{ling99}
J. Shi, X. S. Ling, R. Liang, D. A. Bonn and W. N. Hardy, Phys.
Rev. B {\bf 60}, R12593 (1999);
X.S. Ling, J.E. Berger and D.E. Prober, Phys. Rev. B {\bf 57}, R3249 (1998).

\bibitem{lmv}
A. I. Larkin, M. C. Marchetti and V. M. Vinokur, Phys. Rev. Lett.
{\bf 75}, 2992 (1995).

\bibitem{bh}
S. Bhattacharya and M. J. Higgins, Phys. Rev. Lett. {\bf 70},
2617 (1993); Phys. Rev. B {\bf 49}, 10005 (1994); S.
Bhattacharya, M. J. Higgins and T. V. Ramakrishnan, Phys. Rev.
Lett. {\bf 73}, 1699 (1994).

\bibitem{tang}
C. Tang, X. S. Ling, S. Bhattacharya and P. M. Chaikin, Europhys.
Lett. {\bf 35}, 597 (1996).

\bibitem{ling00}
X. S. Ling, S. R. Park, S. M. Choi, D. C. Dender and J. W. Lynn ,
submitted to Phys. Rev. Lett. (2000).

\bibitem{agy}
T. Ala-Nissila, E. Granato and S. C. Ying, J. Phys.: Condensed
Matt. {\bf 2}, 8537 (1990).

\bibitem{persson}
B. N. J. Persson, {\it Sliding Friction: Physical Principles and
Applications},Springer, Heidelberg, (1998).

\bibitem{robbins}
M. Cieplak, E. D. Smith and M. O. Robbins, Science {\bf 265},
1209 (1994).

\bibitem{tomassone}
M. S. Tomassone, J.B. Sokoloff, A. Widom, and J. Krim, Phys. Rev.
Lett.  {\bf 79}, 4798 (1997).

 \bibitem{gy}
E. Granato and S. C. Ying, Phys. Rev. B {\bf 59}, 5154 (1999).

\bibitem{anderson}
P. W. Anderson, {\it Basic Notions of Condensed Matter Physics},
Addison-Wesley, p.162 (1997).

\bibitem{mori}
H. Mori, Prog. Theor. Phys. {\bf 34}, 399 (1965).

\bibitem{ying}
S. C. Ying, Phys. Rev. B {\bf 41}, 7068 (1989).

\bibitem{halperin69}
B. I. Halperin and P. C. Hohenberg, Phys. Rev. {\bf 177}, 952
(1969).
\bibitem{halperin77}
P. C. Hohenberg and B. I. Halperin, Rev. Mod. Phys. {\bf 49}, 435
(1977).

\bibitem{anomaly}
The possible  critical anomaly of the transport coefficients
depends on whether the damping is due to just the intrinsic
interactions  or an additional linear coupling to external
potentials as in the present case. For details, see Ref.
\onlinecite{halperin77}.

\bibitem{ahy}
T. Ala-Nissila, W. K. Han and S. C. Ying, Phys. Rev. Lett. {\bf
68}, 1866 (1992).

\bibitem{gy00}
M. R. Baldan, E. Granato and S. C. Ying, 
Phys. Rev. B {\bf 62}, 2146 (2000).

\bibitem{young}
J. D. Reger, T. A. Tokuyasu, A. P. Young, and M. P. A. Fisher,
Phys. Rev. B {\bf 44}, 7147 (1991).

\bibitem{giamarchi}
T. Giamarchi and P. Le Doussal, Phys. Rev. B {\bf 52}, 1242
(1995).

\bibitem{nelson}
D. R. Nelson, Phys. Rev. Lett. {\bf 60} , 1973 (1988).

\bibitem{houghton}
A. Houghton, R. A. Pelcovits and A. Sudbo, Phys. Rev. B {\bf 40},
6763 (1989).

\bibitem{huse}
D. S. Fisher, M. P. A. Fisher and D. Huse, Phys. Rev.B {\bf 43},
130 (1991).


\end{thebibliography}
\end{document}